\documentclass{emulateapj}
\usepackage{apjfonts}
\usepackage{mathptmx}

\shorttitle{GAMMA-RAY LUMINOSITY FUNCTION OF BLAZARS}
\shortauthors{Narumoto and Totani}
\slugcomment{Accepted for Publication in The Astrophysical Journal}

\begin{document}

\title{Gamma-Ray Luminosity Function of Blazars and the Cosmic
Gamma-Ray Background: Evidence for the Luminosity Dependent
Density Evolution}

\author{Takuro Narumoto and Tomonori Totani}

\affil{Department of Astronomy, School of Science, Kyoto University,
Sakyo-ku, Kyoto, 606-8502, Japan}

\email{takuro@kusastro.kyoto-u.ac.jp}

\begin{abstract}
We present a comprehensive study for the gamma-ray luminosity function
(GLF) of blazars and their contribution to the extragalactic diffuse
gamma-ray background (EGRB). Radio and gamma-ray luminosity correlation
is introduced with a modest dispersion consistent with observations, to
take into account the radio detectability which is important for the
blazar identification. Previous studies considered only pure luminosity
evolution (PLE) or pure density evolution, but here we introduce the
luminosity dependent density evolution (LDDE) model, which is favored
from the evolution of X-ray luminosity function (XLF) of AGNs. The model
parameters are constrained by likelihood analyses about the observed
redshift and gamma-ray flux distributions of the EGRET blazars.
Interestingly, we find that the LDDE model gives a better fit to the
observed distributions than the PLE model, indicating that the LDDE
model is also appropriate for gamma-ray blazars, and that the jet
activity is universally correlated with the accretion history of AGNs.
The normalization between the GLF and XLF is consistent with the unified
picture of AGNs, when the beaming and a reasonable duty cycle of jet
activity are taken into account. We then find that only $25$--$50\%$ of
the EGRB can be explained by unresolved blazars with the best-fit LDDE
parameters. Unresolved blazars can account for all the EGRB only with a
steeper index of the faint-end slope of the GLF, which is marginally
consistent with the EGRET data but inconsistent with that of the XLF.
Therefore unresolved AGNs cannot be the dominant source of the EGRB,
unless there is a new population of gamma-ray emitting AGNs that evolves
differently from the XLF of AGNs. Predictions for the \textit{GLAST}
mission are made, and we find that the best-fit LDDE model predicts
about 3000 blazars in the entire sky, which is considerably fewer (by a
factor of more than three) than a previous estimate.
\end{abstract}


\section{Introduction}
The origin of the extragalactic diffuse gamma-ray background (EGRB) is
one of the unsolved problems in astrophysics.  The EGRB was first
discovered by the \textit{SAS} 2 satellite (Fichtel, Simpson, \& Thompson
1978; Thompson \& Fichtel 1982) and subsequently confirmed by the
Energetic Gamma Ray Experiment Telescope (EGRET) instrument aboard the
\textit{Compton Gamma Ray Observatory (CGRO)}. In the first analysis of
the EGRET data, the flux of the EGRB integrated above 100 MeV was
determined to be $(1.45\pm0.05)\times10^{-5}$ photons $\rm{cm}^{-2}$
$\rm{s}^{-1}$ $\rm{sr}^{-1}$ (Sreekumar et al. 1998). However, this value
is strongly dependent on the modeling of the Galactic background, which
is dominated by cosmic-ray interactions in the Galactic disk and must be
subtracted from the background data. The latest analysis, which used a new
model of the Galactic background, resulted in a slightly smaller value of
the EGRB, $(1.14\pm0.12)\times10^{-5}$ photons $\rm{cm}^{-2}$ $\rm{s}^{-1}$
$\rm{sr}^{-1}$ (Strong et al. 2004).

EGRET detected many extragalactic high-energy gamma-ray sources that have
been identified as active galactic nuclei (AGNs). Most of them fall into
the blazar class of AGNs, and this is the only one extragalactic
population confirmed in the third EGRET catalog (Hartman et al. 1999),
constituting about 15\% of the EGRB flux. Therefore unresolved blazars
are the most likely candidate for the origin of the EGRB, at least
contributing substantially, and this issue has been studied in a number
of papers (Padovani et al. 1993; Stecker, Salamon, \& Malkan 1993;
Salamon \& Stecker 1994; Chiang et al. 1995; Stecker \& Salamon 1996;
Chiang \& Mukherjee 1998; M\"ucke \& Pohl 2000). On the other hand,
several alternative candidates for the EGRB components have been proposed,
e.g., intergalactic shocks produced by the formation of large-scale
cosmological structures (Loeb \& Waxman 2000; Totani \& Kitayama 2000;
Miniati 2002; Scharf \& Mukherjee 2002; Gabici \& Blasi 2003; Keshet et
al. 2003), or dark matter annihilation (Oda, Totani, \& Nagashima 2005
and references therein). Therefore, it is important to determine whether
the number of unresolved blazars are enough to account for all of the EGRB,
but conclusions derived by the earlier studies are somewhat controversial.

Stecker \& Salamon (1996, hereafter SS96) estimated the unresolved blazar
contribution with basic assumptions that EGRET blazars are the same
population with flat-spectrum radio-loud quasars (FSRQs), and that the
gamma-ray and radio luminosities are linearly related. Then they
constructed the blazar gamma-ray luminosity function (GLF) model from
the FSRQ radio luminosity function (RLF), and found that blazars can
account for 100\% of the EGRB. However, their model was not compared
with the available redshift distribution of the EGRET blazars, and
hence it was uncertain whether this GLF model is statistically
consistent with the EGRET blazar data.

Quantitative comparison of GLF models to the flux and redshift
distributions of the EGRET blazars was performed by Chiang \& Mukherjee
(1998, hereafter CM98), and indeed, they found that the model of SS96
seriously overpredicts the number of low-redshift blazars detectable by
the EGRET. CM98 then concluded that blazars can account for only 25\% of
the EGRB, based on the GLF model consistent with the EGRET blazar
distributions.\footnote{In this paper, we refer the fraction of the blazar
contribution in the EGRB always removing the already detected EGRET blazars,
i.e., the ratio of the background flux from blazars under the EGRET
detection limit to the EGRB flux of $1.14\times10^{-5}$ photons
$\rm{cm}^{-2}$ $\rm{s}^{-1}$ reported by Strong et al. (2004).}

However, the analysis of the GLF is not straightforward; a source of
uncertainty is the detectability in the radio band. Most of the EGRET
blazars are identified by finding radio counterparts, and hence they
would remain unidentified if their radio counterparts are under the flux
limit of radio surveys, even though their gamma-ray flux is above the
EGRET sensitivity limit. Therefore one must estimate the probability of
a model blazar having flux greater than the sensitivity limits, not only
in the gamma-ray band but also in the radio band.  CM98 introduced this
probability in their analysis, but they assumed that there is no
correlation between gamma-ray and radio luminosities of blazars. However,
the assumption of no correlation at all over a wide range of gamma-ray
and radio luminosities induces some inconsistencies (see discussion given
in Stecker \& Salamon 2001), and it is physically reasonable to expect
some level of correlation from the viewpoint of the standard
synchrotron-inverse Compton model of blazars. Therefore we adopt a new
treatment on this issue introducing a reasonable correlation (see
\S\ref{section:radio-ID} for details).

M\"ucke \& Pohl (2000) approached this issue from the viewpoint of the
unification scheme of radio-loud AGNs, which proposed that blazars are
the beamed subclass of Fanaroff-Riley (FR) radio galaxies. They considered
the GLF models based on the RLF of FR galaxies, and the flux and redshift
distributions of blazars were used to constrain their GLF models. Then
they concluded that unresolved blazar contributuion to the EGRB is 20--40\%
assuming that the blazars extend to the maximum cutoff redshift of
$z_{\rm{max}}=3$, and 40--80\% for $z_{\rm{max}}=5$. However, the
identification probability of a blazar, which may affect the estimate of
the blazar contribution to the EGRB, was not incorporated in their analysis.

To resolve this rather controversial situation, in this paper we make a
comprehensive study of GLF models that are statistically compared with
the observed redshift and flux distributions, taking into account a
reasonable correlation between gamma-ray and radio flux in a consistent
manner with the observed gamma-ray to radio flux ratios. Then we make an
estimate of the blazar contribution to the EGRB flux, and also estimate
the expected number of ``unidentified blazars'' due to the lack of radio
detection, which can be compared with the number of high-Galactic-latitude
unidentified sources in the third EGRET catalog. We also make some
predictions for the future \textit{Gamma Ray Large Area Space Telescope}
(\textit{GLAST}) observation, and discuss its prospects.

In addition to these new aspects, we also try a new type of the GLF
evolution model. The earlier studies treated the cosmological evolution
of the blazar GLF as a pure luminosity evolution (PLE) or a pure density
evolution. On the other hand, the cosmological evolution of the luminosity
function of AGNs has been investigated intensively in soft X-ray (e.g.,
Miyaji, Hasinger, \& Schmidt 2000, 2001; Hasinger, Miyaji, \& Schmidt 2005)
and hard X-ray (e.g., Boyle et al. 1998; La Franca et al. 2002; Cowie et al.
2003; Ueda et al. 2003) bands, recently. The former studies are mostly for type
1 AGNs, while the latter ones are for both type 1 and 2 AGNs, since soft X-ray
emission of type 2 AGNs is significantly absorbed. These studies revealed
that the overall behavior of the soft X-ray luminosity function (SXLF)
and hard X-ray luminosity function (HXLF) of AGNs are very similar and best
described with a luminosity dependent density evolution (LDDE) where the
peak redshift of density evolution increases with AGN luminosity (Miyaji,
Hasinger, \& Schmidt 2000, 2001; Ueda et al. 2003; Hasinger, Miyaji, \&
Schmidt 2005). Therefore it is reasonable to expect that the cosmological
evolution of the blazar GLF may also be expressed by the LDDE. In this
paper, we try two kinds of blazar GLF model; one is based on the FSRQ RLF
(PLE model) and the other is based on the AGN SXLF (LDDE model).

The paper will be organized as follows: In \S2, we describe sample
definition and formulations for the statistical analyses. In \S3, we
present models of the blazar GLF and results of comparison to the EGRET
data. In \S4, we address the prospects for the \textit{GLAST} mission.
Discussions and conclusions are given in \S5 and \S6, respectively.
Throughout this paper, we adopt a $\Lambda$CDM universe with the density
parameter $\Omega_{0}=0.3$, the cosmological constant $\Omega_{\Lambda}=0.7$,
and the Hubble constant $H_0=70$ $\rm{km}$ $\rm{s}^{-1}$ $\rm{Mpc}^{-3}$.

\section{Formulations}
\subsection{Sample Definition}
Most of blazars show considerable variability, but it is difficult to
incorporate in the statistical analysis of the GLF. Therefore we use
the mean flux shown in the third EGRET catalog. There are some blazars
in the EGRET catalog whose significance of detection is above the EGRET
threshold ($4\sigma$ for $|b|>10^\circ$, and $5\sigma$ for $|b|<10^\circ$)
only in some viewing periods, and their mean significance of detection is
under the EGRET threshold. We exclude such blazars from our analysis, and
use only blazars whose mean significance of detection (as shown in the
catalog) is above the EGRET threshold. There are 46 blazars meeting this
selection.

\begin{figure*}
\epsscale{1.1}
\plotone{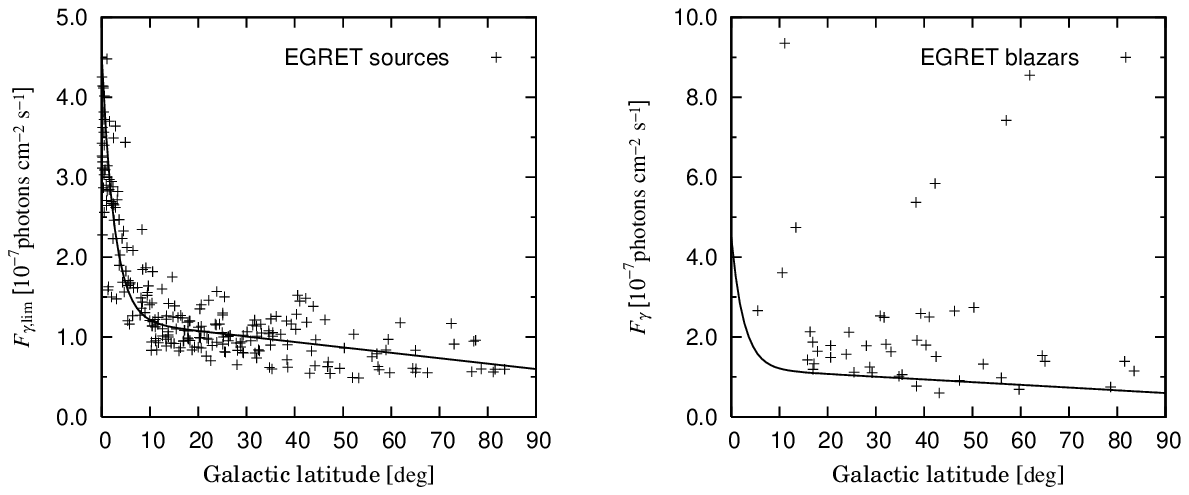}
\caption{Left panel: Sensitivity limit to point sources as a function
of the Galactic latitude in the third EGRET catalog. The fitted relation
is also shown by the solid line. Right panel: mean flux of the EGRET
blazars against its Galactic latitude. The solid line is the same as
the left panel.}
\label{fig:F-latitude}
\end{figure*}

Since the detection limit is sensitively dependent on the location in
the sky, we will take it into account in the analysis. In the left panel
of Figure \ref{fig:F-latitude} we plot the EGRET detection limit for each
EGRET point source, which is calculated from mean flux and mean statistical
significance of the detection by using equation (1) of CM98, against its
Galactic latitude (\textit{crosses}). The best-fit relation is also shown
(\textit{solid line}). Because of the variability, this fit may be different
from the effective detection threshold for blazars that we have chosen. In
order to check this point, in the right panel of Figure \ref{fig:F-latitude}
we plot the mean flux of the EGRET blazars, comparing them to the obtained
detection limit curve. We see from this figure that most of blazars are
above the detection limit curve, and hence it is reasonable to take this
curve as the EGRET detection limit for the 46 blazars used in this paper.

It is difficult to take into account the variety of blazar spectra, and
we assume a single universal power-law spectrum for all blazars, with
the photon spectral index of $\alpha=2.2$
($dN_{\gamma}/dE_{\gamma}\propto E_{\gamma}^{-\alpha}$). This value is
close to the mean of the EGRET blazars ($\alpha=2.15\pm0.04$, Sreekumar
et al. 1998). We confirmed that our conclusions in this paper are not
seriously affected even if we change the value of this parameter into
$\alpha=2.1$ and $2.3$. Then, the gamma-ray luminosity $L_{\gamma}$,
which is defined as $\nu L_{\nu}$ in erg $\rm{s}^{-1}$ at 100 MeV (at
the restframe), is related to the observed photon flux, $F_{\gamma}$
at $E_{\gamma}\geq E_{\min}\equiv$ 100 MeV (in photons $\rm{cm}^{-2}$
$\rm{s}^{-1}$ and $E_{\min}$ in the observer's frame), as
\begin{eqnarray}
L_{\gamma}=4\pi d_L^2\ \frac{\alpha-1}{(1+z)^{2-\alpha}}\ E_{\min}\
F_\gamma\ ,
\label{eq:Luminosity-Flux}
\end{eqnarray}
where $d_L$ is the standard luminosity distance. Then, the observed
data that will be used in the statistical analysis is a set of
$(z_i,L_{\gamma,i},\Omega_i)$, where $\Omega_i$ denotes the observed
blazar location in the sky, and the subscript $i$ denotes each blazar,
running over $1\leq i\leq N_{\rm{obs}}=46$.

\subsection{Maximum Likelihood Method}
We constrain the model parameters of the blazar GLF models by using
the maximum likelihood method. A specified GLF model predicts the
distribution function of the three quantities,
$d^3N/(dz\ dL_{\gamma}\ d\Omega)$, and the likelihood function
${\cal L}$ is given as (see, e.g., Loredo \& Lamb 1989):
\begin{equation}
{\cal L}=\exp(-N_{\rm{exp}})\ \prod_{i=1}^{N_{\rm{obs}}}\ 
\frac{d^3N(z_i,L_{\gamma,i},\Omega_i)}{dz\ dL_{\gamma}\ d\Omega}\ , 
\end{equation}
where $N_{\rm{exp}}$ is the expected number of blazar detections:
\begin{equation}
N_{\rm{exp}}=\int dz\int dL_{\gamma}\int
d\Omega\ \frac{d^3N}{dz\ dL_{\gamma}\ d\Omega}\ .
\end{equation}
Consider a transformation about the normalization,
$d^3N/(dz\ dL_{\gamma}\ d\Omega)\rightarrow fd^3N/(dz\ dL_{\gamma}\ d\Omega)$.
By maximizing the likelihood function about $f$, we find
$f=N_{\rm{obs}}/N_{\rm{exp}}$, i.e., the maximum likelihood obtained when
the expected number, $fN_{\rm{exp}}$, becomes equal to $N_{\rm{obs}}$.
Substituting $f=N_{\rm{obs}}/N_{\rm{exp}}$ and ignoring constant factors
that are not relevant for the likelihood maximization, we find the
normalization-free form of the likelihood function:
\begin{equation}
{\cal L}=\prod_{i=1}^{N_{\rm{obs}}}\ \left(\frac{1}{N_{\rm{exp}}}\
\frac{d^3N(z_i,L_{\gamma,i},\Omega_i)}{dz\ dL_{\gamma}\ d\Omega}\right)\ . 
\end{equation}
The distribution function can be expressed as
\begin{eqnarray}
\frac{d^3N}{dz\ dL_{\gamma}\ d\Omega}&=&\frac{dV}{dz}\
\rho_{\gamma}(L_{\gamma},z)\ \epsilon(L_\gamma,z)\nonumber\\
&\times&\Theta\left[F_{\gamma}(L_{\gamma},z)-F_{\gamma,\rm{lim}}
(\Omega)\right]\ ,
\end{eqnarray}
where $\rho_{\gamma}$ is the GLF per unit comoving density and unit
luminosity, $dV/dz$ is the comoving volume element per unit solid
angle as defined in the standard cosmology, and $\Theta$ is the step
function [$\Theta(x)=1$ and 0 for $x\geq0$ and $<0$, respectively].
The detection efficiency $\epsilon(L_{\gamma},z)$ represents the
identification probability as a blazar by finding a radio counterpart,
which will be defined in the next subsection.

To find the best-fit model parameters and their confidence regions, we
use the standard likelihood ratio method, assuming that
${\cal L}\propto\exp(-\chi^2/2)$, where $\chi^2$ obeys the chi-square
distribution. The best-fit parameters are simply obtained as those
giving the minimum chi-square, $\chi_{\min}^2$, and the confidence
region is determined by the contour of a constant
$\Delta\chi^2\equiv\chi^2-\chi_{\min}^2$. In this paper we will perform
two-parameter fit to the data, and hence $\Delta\chi^2$ obeys to the
chi-square distribution with two degrees of freedom, i.e.,
$\Delta\chi^2=2.30$, $6.16$, and $9.21$ for 68\%, 95\%, and 99\% C.L.,
respectively (see, e.g., Press et al. 1992).

\subsection{Identification Probability in Radio Band}
\label{section:radio-ID}
\begin{figure*}
\epsscale{1.1}
\plotone{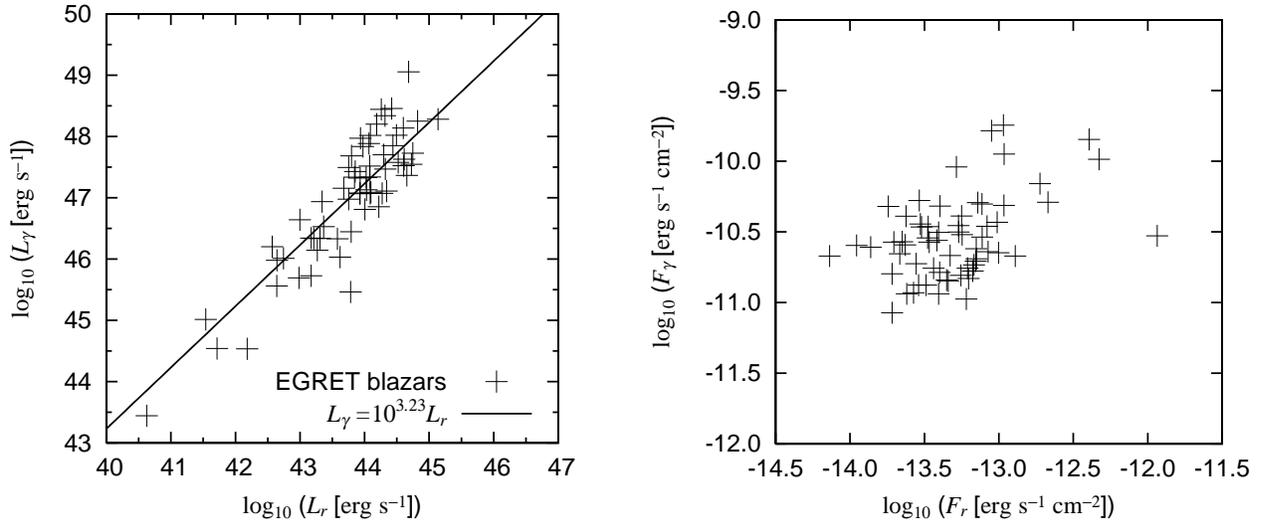}
\caption{Left (Right) panel: Observed gamma-ray and radio luminosities
(fluxes) of the EGRET blazars are shown by the crosses. Here, the luminosity
(flux) is defined as $\nu L_{\nu}$ ($\nu F_{\nu}$) at restframe (observed)
100 MeV and 2.7 GHz for the gamma-ray and radio bands, respectively. The
solid line is the best-fit linear relation. The K-corrections are done for
luminosities assuming typical spectral indices (see text).}
\label{fig:gamma-radio}
\end{figure*}

Here we formulate the probability that a gamma-ray blazar having a
gamma-ray flux above the EGRET detection limit is also detected
in the radio band, so that it is identified as an EGRET blazar. 
In the left panel of Figure \ref{fig:gamma-radio} we show the
observed relation between $L_\gamma$ and radio luminosity $L_r$
($\nu L_{\nu}$ in erg $\rm{s}^{-1}$ at restframe 2.7 GHz) of the
EGRET blazars. The best-fit relation is also shown in the figure
(\textit{solid line}). Here we assumed the photon spectral index
of $\alpha_r=1.0$ for the K-correction, as a typical index of blazars
in the radio band (M\"{u}cke et al. 1997). It should be noted that,
though the gamma-ray and radio luminosities are apparently well
correlated with each other, this is mostly an artifact, coming from
the fact that blazars with different distances are detected with a
similar flux around the detection limit. This can easily be understood
if we see the correlation plot between observed gamma-ray and radio
fluxes, as shown in the right panel of the same figure.

The correlation between gamma-ray and radio emissions of blazars
has been investigated and the evidence for this correlation has
been presented in many papers (e.g., Padovani et al. 1993; Stecker,
Salamon, \& Malkan 1993; Salamon \& Stecker 1994; Dondi \& Ghisellini
1995; L\"{a}hteenm\"{a}ki et al. 1997; L\"{a}hteenm\"{a}ki, Valtaoja,
\& Tornikoski 2000; Tornikoski \& L\"{a}hteenm\"{a}ki 2000;
L\"{a}hteenm\"{a}ki \& Valtaoja 2001). On the other hand, M\"{u}cke et
al. (1997) claimed that correlation between the gamma-ray and radio
luminosities cannot be established firmly from the existing data. This
is probably because the correlation is hidden by intrinsic dispersion
and the rather narrow dynamic range of observed radio and gamma-ray
fluxes. Based on this result, CM98 assumed no correlation between
$L_\gamma$ and $L_r$.

However, it should be noted that the dynamic range of luminosity of
EGRET blazars is extending over more than five order of magnitudes
(Fig. \ref{fig:gamma-radio}). If you assume no correlation between
gamma-ray and radio luminosities, it means that you cannot tell which
blazar is brighter in the radio band, even if you already know that one
blazar is brighter than the other by a factor of $10^5$. Such an
assumption is highly unlikely, since it is generally believed that the
overall spectra of blazars are made by two different emission processes
from the same population of relativistic electrons; the gamma-ray
emission is due to the inverse Compton process, while the radio emission
is due to the synchrotron process. Therefore, we must introduce some
correlation between the gamma-ray and radio luminosities in the analysis.

\begin{figure}
\epsscale{0.55}
\plotone{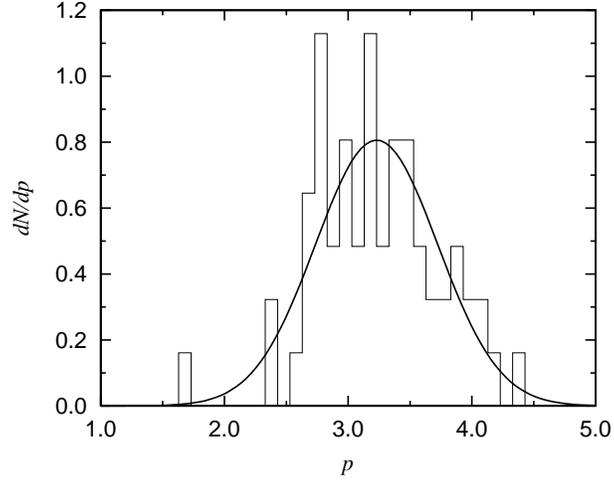}
\caption{Histogram of radio to gamma-ray luminosity ratio,
$p=\log_{10}(L_{\gamma}/L_r)$, of the EGRET blazars. The luminosities
are $\nu L_{\nu}$ in the restframe 100 MeV and 2.7 GHz bands, respectively.
The solid curve is a Gaussian fit to the histogram.}
\label{fig:G-R_scatter}
\end{figure}

Hence we introduce a linear correlation with log-normal scatter as a
simple and phenomenological model, to avoid theoretical uncertainties of
more physically motivated models. Then, $L_{\gamma}/L_r$ obeys to the
log-normal distribution with $\langle p\rangle\sim3.23$, where
$p\equiv\log_{10}(L_{\gamma}/L_r)$. Figure \ref{fig:G-R_scatter} shows
the distribution of $p$, and the best-fit dispersion is $\sigma_p=0.49$.
Then, the probability that a blazar having gamma-ray luminosity
$L_{\gamma}$ at redshift $z$ will be identified in radio band can be
calculated as the probability that the corresponding radio flux at 2.7
GHz (at observer's frame) is greater than the radio detection limit,
$F_{r,\lim}$. We take $F_{r,\lim}=0.7$ Jy, since most of the EGRET
blazars have radio fluxes larger than 0.7 Jy.

It should be noted that this correlation may be different from the true
correlation between $L_{\gamma}$ and $L_r$, since the observed correlation
has been affected by selection effects. The ratio of the flux limits in
radio and gamma-ray bands (0.7 Jy at 2.7 GHz and $\sim1.0\times 10^{-7}$
photons $\rm{cm}^{-2}$ $\rm{s}^{-1}$ above 100 MeV, respectively) is in
fact very close to the mean of $p$. In order to check how much our
analysis could be affected by this effect, we calculated the prediction
of the $p$ distribution that will actually be observed for the EGRET
blazars, from the best-fit models of the blazar GLF that will be obtained
later in this paper. We confirmed that the predicted distribution is
consistent with the observed one, and this consistency check demonstrates
that our analysis is not seriously biased by the selection effect.

Though we assumed a simple linear relation, more physically motivated
models such as the synchrotron self-Compton (SSC) model and the
external radiation Compton (ERC) model may predict deviation from the
exact linear correlation. However, the above result indicates that the
linear correlation is statistically consistent with the data, and we
cannot derive more detailed conclusions from the current sample. The
future \textit{GLAST} mission may provide better statistics for this
issue. We will discuss about the SSC and ERC models in the context of
the beaming factor difference in the gamma-ray and radio bands in \S
\ref{section:beaming}.

\subsection{Background Photon Flux from Unresolved Blazars}
We can calculate the integrated background photon flux ($>100$ MeV) from
blazars below the EGRET detection limit as
\begin{eqnarray}
F_{\rm{diffuse}}=\int^{z_{\rm{max}}}_{0}dz\ \frac{dV}{dz}
\int^{L_{\gamma,\rm{lim}}(z)}_{L_{\gamma,\rm{min}}}dL_{\gamma}\ 
F_{\gamma}(L_{\gamma},z)\ \rho_{\gamma}(L_{\gamma},z)\ ,\quad
\end{eqnarray}
where $L_{\gamma,\rm{lim}}(z)$ is the gamma-ray luminosity corresponding
to the EGRET threshold, and $L_{\gamma,\rm{min}}$ is the minimum gamma-ray
luminosity of the blazar GLF. This quantity will be compared with the
observed EGRB, to estimate the contribution from unresolved blazars.
Since the minimum gamma-ray luminosity is quite uncertain and has
considerable effect on the blazar contribution to the EGRB, we consider
four values of $L_{\gamma,\rm{min}}=10^{43}$, $10^{42}$, $10^{41}$, and
$10^{40}$ erg $\rm{s}^{-1}$. For reference, $L_{\gamma,\rm{min}}=10^{43}$
erg $\rm{s}^{-1}$ is smaller than the minimum gamma-ray luminosity of the
EGRET blazars by a factor of $\sim5$. We assume $z_{\max}=5$, but the
predicted EGRB flux hardly depends on this parameter, since the number
density of AGNs with a given luminosity decreases with redshift beyond
$z\sim2$ by the estimated evolution of luminosity functions in the radio
or X-ray bands, based on which our GLF models will be constructed.

\section{Models of the Blazar Gamma-Ray Luminosity Function
and Results of the Analysis}
In this paper we try two models of the blazar GLF,
$\rho_{\gamma}(L_{\gamma},z)$. The descriptions of these two models
and fits to the observed data will be presented below.

\subsection{The Pure Luminosity Evolution (PLE) Model}
\subsubsection{Model Description}
\label{section:PLE-description}
For the PLE model, we follow the same procedure proposed by SS96
for constructing the blazar GLF model. They made the basic assumptions
that blazars seen by gamma-rays above 100 MeV are also seen in radio
as FSRQs, and that the gamma-ray and radio luminosities of these objects
are linearly related as
\begin{equation}
L_{\gamma}= 10^{\langle p\rangle}L_r\ ,
\label{eq:gamma-radio}
\end{equation}
where the definitions and units are the same as defined in the previous
section. The blazar GLF is then derived from the FSRQ RLF:
\begin{equation}
\rho_{\gamma}(L_{\gamma},z)=\eta\ \frac{L_r}{L_{\gamma}}\ \rho_{r}(L_r,z)\ ,
\end{equation}
where $\eta$ is a normalization factor, and $\rho_{r}(L_{r},z)$
is the FSRQ RLF. We use the FSRQ RLF derived by Dunlop \& Peacock
(1990, hereafter DP90):
\begin{equation}
\rho_r(L_r,z)=\frac{1}{f(z)}\ \rho_r\left(\frac{L_r}{f(z)},0\right)\ ,
\end{equation}
where $\rho_r(L_r,0)$ is the present-day FSRQ RLF, which is characterized
by the faint-end slope index $\gamma_1$, the bright-end slope index $\gamma_2$,
and the break luminosity $L^{\ast}_r$, given as
\begin{equation}
\rho_{r}(L_{r},0)=\frac{A_r}{(\ln10)\ L_r}
\left\{\left[\frac{L_r}{L^{\ast}_r}\right]^{\gamma_1}+
\left[\frac{L_r}{L^{\ast}_r}\right]^{\gamma_2}\right\}^{-1},
\end{equation}
and $f(z)$ is the luminosity evolution function given as
\begin{equation}
f(z)=10^{az+bz^2} \ .
\end{equation}
Here, $A_r=7.08\times10^{-9}$ $\rm{Mpc}^{-3}$, $\log_{10}L^{\ast}_r=42.79$,
$\gamma_1=0.83$, $\gamma_2=1.96$, $a=1.18$, and $b=-0.28$. Since this FSRQ
RLF was derived for the Einstein-de Sitter (EdS) universe with
$(\Omega_0,\Omega_\Lambda)=(1.0,0.0)$ and $H_0=50$ $\rm{km}$
$\rm{s}^{-1}$ $\rm{Mpc}^{-3}$, we multiply $\rho_r(L_r,z)$ by a correction
factor $\textstyle(dV_{\rm{EdS}}/dV_{\rm{\Lambda}})$ and $f(z)$ by a
correction factor $\textstyle(d_{L,\rm{\Lambda}}/d_{L,\rm{EdS}})^2$ in
order to approximately convert to a FSRQ RLF for the $\Lambda$CDM universe,
where $dV$ is the comoving volume element of the universe, and $d_L$
is the luminosity distance. Therefore, this model is no longer ``PLE''
model in a strict sense, but we take this correction into account to apply
the RLF in agreement with the observed data.

\subsubsection{Constraints from the Redshift and Luminosity
Distribution of the EGRET blazars}
\begin{figure}
\epsscale{0.55}
\plotone{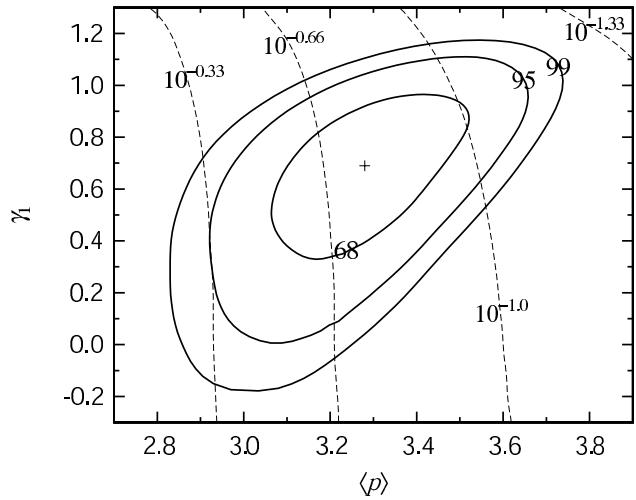}
\caption{The solid contours show the 68\%, 95\% and 99\% C.L. regions
for the PLE model parameters [the faint-end slope index $\gamma_1$ and
the mean gamma-ray to radio luminosity ratio, 
$\langle p\rangle=\langle\log_{10}(L_{\gamma}/L_r)\rangle$].
The best-fit values, $(\langle p\rangle,\gamma_1)=(3.28,0.69)$,
are shown by the cross. The dashed contours correspond to $\eta=10^{-0.33}$,
$10^{-0.66}$, $10^{-1.0}$, and $10^{-1.33}$, respectively, where $\eta$
is the ratio of the normalizations of the gamma-ray to radio luminosity
functions.}
\label{fig:CL_contour_PLE}
\end{figure}

In this model, we take $\langle p\rangle$ and $\gamma_1$ as the two
free parameters since they are poorly constrained from observations,
and fix the other parameters to the best-fit values shown in
\S\ref{section:PLE-description}. For consistency, we use the same
$\langle p\rangle$ with the dispersion $\sigma_p=0.49$ also for
judgement of the radio identification, which has been described in
\S\ref{section:radio-ID}. The normalization factor $\eta$, which is
physically related to the possible beaming effect, is determined by the
requirement that the calculated number of identifiable blazars above
the EGRET threshold is equivalent to the observed number of the EGRET 
blazars. In Figure \ref{fig:CL_contour_PLE} we show the 68\%, 95\%, and
99\% C.L. contours for the PLE model parameters (\textit{solid lines}).
The best-fit parameters, $(\langle p\rangle,\gamma_1)=(3.28,0.69)$, are
also marked (\textit{cross}). The value $\langle p\rangle=3.28$ is quite
similar to the value directly obtained from the EGRET blazars ($p=3.23$).
The faint end slope $\gamma_1=0.69$ is somewhat smaller (i.e., flatter
faint-end slope) than that of the FSRQ RLF derived by DP90 $(\gamma_1=0.83)$,
but the value of DP90 is well within the 68\% C.L. region. 

\begin{figure}
\epsscale{0.55}
\plotone{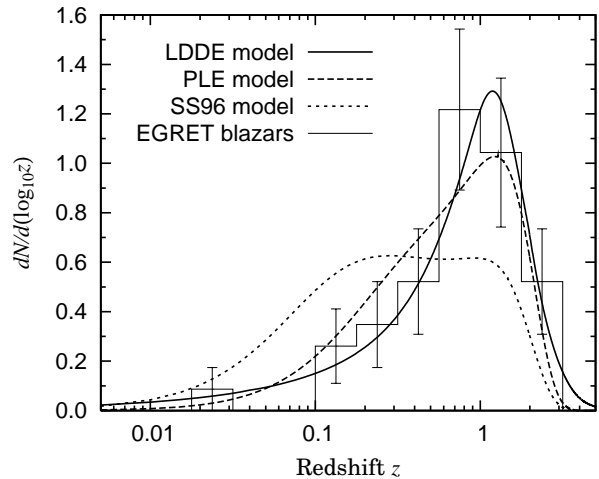}
\caption{Redshift distribution of the EGRET blazars. The histogram is
the EGRET data. The solid and dashed curves are the best-fit models for
the LDDE and PLE models, respectively, from the likelihood analysis. The
dotted curve is obtained from the blazar GLF model of SS96. The error
bars are 1$\sigma$ Poisson error.}
\label{fig:z_dist_EGRET}
\end{figure}

\begin{figure}
\epsscale{0.55}
\plotone{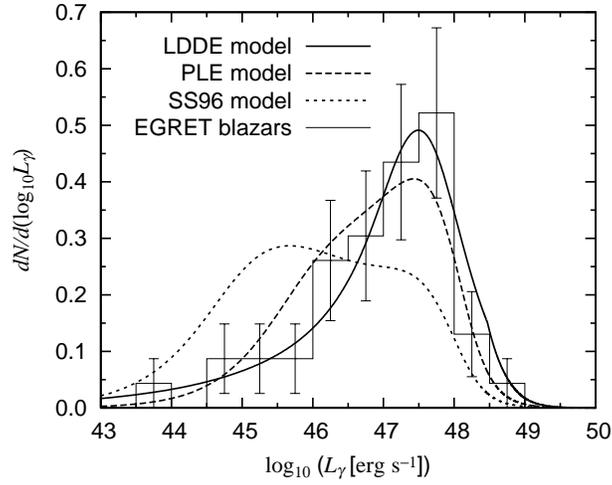}
\caption{Luminosity distribution of the EGRET blazars. The line
markings are the same as Figure \ref{fig:z_dist_EGRET}. The luminosity
is $\nu L_{\nu}$ at 100 MeV. The error bars are 1$\sigma$ Poisson error.}
\label{fig:L_dist_EGRET}
\end{figure}

Figures \ref{fig:z_dist_EGRET} and \ref{fig:L_dist_EGRET} show the
redshift and luminosity distributions for the best-fit parameters,
respectively (\textit{dashed lines}). It is clear that the PLE model
with parameters adopted by SS96 ($\langle p\rangle=2.54$ and
$\gamma_1=0.83$ from DP90) can reproduce neither the redshift nor
luminosity distributions. Our best-fit model reproduces these
distributions better than the SS96 model, but still the fit is
not very good, especially for the redshift distribution. We performed
the Kolmogorov-Smirnov (KS) test, and find that the chance probability
of getting the observed deviation of the redshift distribution from the
best-fit PLE model is only 3.1\%, while it is 27.0\% for the luminosity
distribution. These results indicate that the PLE framework may {\it not}
be satisfactory to describe the EGRET blazar data.

\subsubsection{Blazar Contribution to the Extragalactic Diffuse
Gamma-Ray Background}
\begin{figure*}
\epsscale{1.1}
\plotone{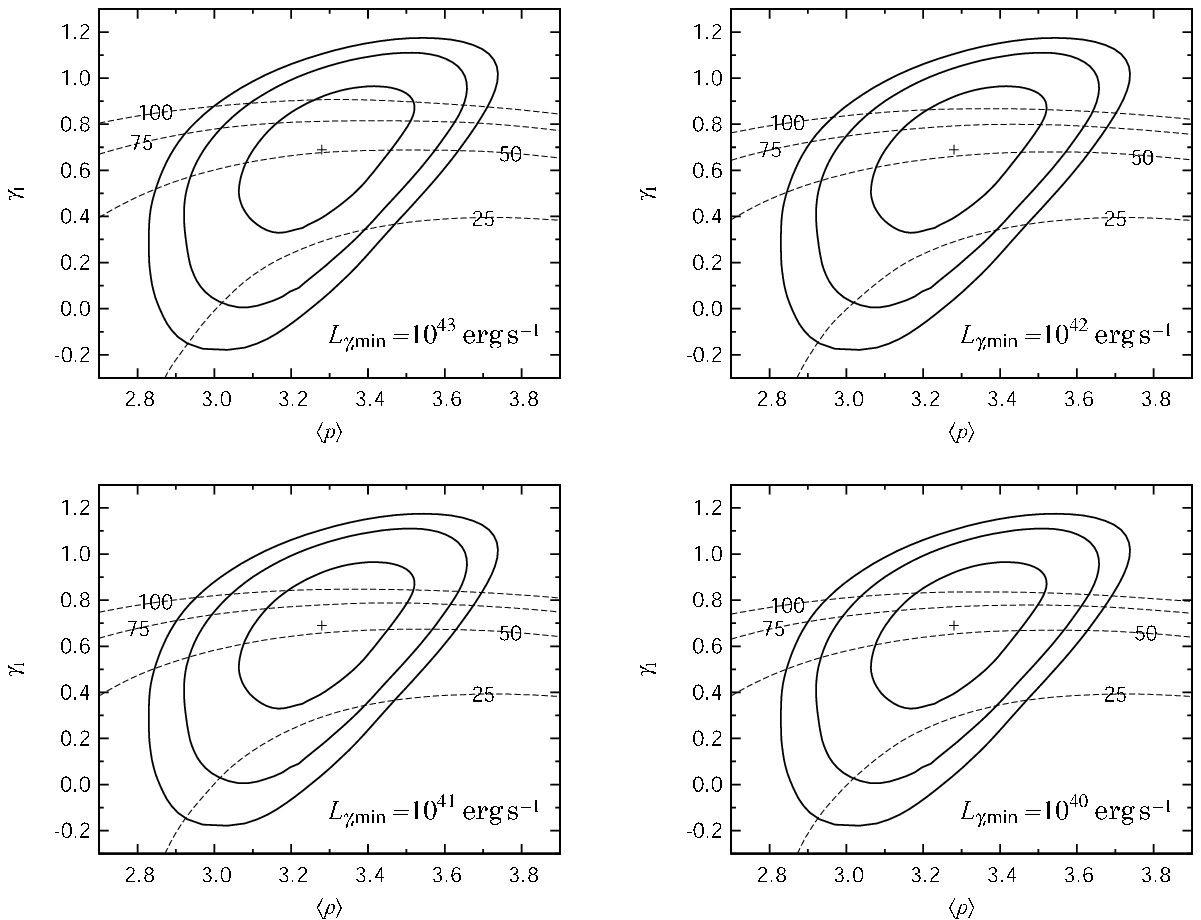}
\caption{The solid contours and crosses are the same as Figure
\ref{fig:CL_contour_PLE} showing the fit by the PLE model. The dashed
contours show 25\%, 50\%, 75\%, and 100\% contribution of unresolved
blazars to the EGRB. The upper left, upper right, lower left, and lower
right panels are for the cases of $L_{\gamma,\rm{min}}=10^{43}$, $10^{42}$,
$10^{41}$, and $10^{40}$ erg $\rm{s}^{-1}$, respectively.}
\label{fig:EGRB_PLE}
\end{figure*}

In Figure \ref{fig:EGRB_PLE} we present the contours of 25\%, 50\%, 75\%,
and 100\% contribution of blazars under the EGRET sensitivity limit to
the EGRB for the PLE model (\textit{dashed lines}). We find that unresolved
blazars can explain only 50--55\% of the EGRB for the best-fit parameters.
On the other hand, since the contour of 100\% blazar contribution pass
through inside the 68\% C.L. region for all the cases, it is unable to
exclude the possibility that almost all of the EGRB is explained by blazars.
However, the poor fit of the PLE model to the observed redshift distribution
indicates that it is not appropriate to derive any conclusion about the EGRB
based on this model framework. It is apparent from this figure that the
blazar contribution to the EGRB is strongly dependent on the faint end slope
$\gamma_1$. On the other hand, since the EGRB contribution becomes 100\% in
a region where $\gamma_1<1.0$, the minimum gamma-ray luminosity of blazars,
$L_{\gamma,\rm{min}}$, hardly affects the contribution to the EGRB.

\subsection{The Luminosity Dependent Density Evolution (LDDE) Model}
\subsubsection{Model Description}
In this section, we construct the blazar GLF model based on the AGN SXLF
by assuming a linear relation between the blazar gamma-ray luminosity
(dominated by the jet) and the AGN soft X-ray luminosity (dominated by
the disk emission) expressed as
\begin{equation}
L_{\gamma}=10^{q}L_{X}\ ,
\end{equation}
where the unit of $L_{\gamma}$ (in $\nu L_{\nu}$ at the restframe 100 MeV)
and $L_{X}$ (in the restframe 0.5--2 keV X-ray band) is erg $\rm{s}^{-1}$.
In the soft X-ray bands, the typical AGN spectra have a photon index of
$\sim2$, i.e., constant in $\nu F_{\nu}$, and hence $L_X$ in the observed
0.5--2 keV band is used as that in the restframe 0.5--2 keV band (e.g.,
Hasinger, Miyaji \& Schmidt 2005). The assumption of the linear relation
between $L_X$ and $L_{\gamma}$ is motivated from an expectation that the
jet activity should be somehow correlated with the accretion power which
can be measured by X-ray luminosity from accretion disks. It should be noted
that this relation is not necessarily be seen in the observed spectral
energy distributions of blazars, since X-ray emission from blazars is
dominated by the beamed emission from the jet, rather than the disk emission.

The blazar GLF is then obtained from the AGN SXLF:
\begin{equation}
\rho_{\gamma}(L_{\gamma},z)=\kappa\ \frac{L_X}{L_\gamma}\ \rho_{X}(L_X,z)\ ,
\end{equation}
where $\kappa$ is a normalization factor, and $\rho_X(L_X,z)$ is the
AGN SXLF. In this model, we adopt the same form as derived by
Hasinger, Miyaji, \& Schmidt (2005) for the AGN SXLF, and the details are
as follows:
\begin{equation}
\rho_X(L_X,z)=\rho_X(L_X,0)\ f(L_X,z)\ ,
\end{equation}
where $\rho_X(L_X,0)$ is the present-day AGN SXLF, which is characterized
by the faint-end slope index $\gamma_1$, the bright-end slope index $\gamma_2$,
and the break luminosity $L^{\ast}_X$, given as
\begin{equation}
\rho_X(L_X,0)=\frac{A_X}{(\ln 10)\ L_X}\left\{\left[\frac{L_X}{L^{\ast}_X}
\right]^{\gamma_1}+\left[\frac{L_X}{L^{\ast}_X}\right]^{\gamma_2}\right\}^{-1},
\end{equation}
and $f(L_X,z)$ is the density evolution function given as
\begin{equation}
f(L_X,z)=\left\{
  \begin{array}{ll}
    (1+z)^{p_1}&\left[z\leq z_c(L_X)\right]\ ,\\
    f\left[L_X, z_c(L_X)\right]\left[\displaystyle{\frac{1+z}{1+z_c(L_X)}}
    \right]^{p_2}&\left[z>z_c(L_X)\right]\ ,\\
  \end{array}
\right.
\end{equation}
where $z_c$ is the redshift of evolutionary peak given as
\begin{equation}
z_c(L_X)=\left\{
  \begin{array}{ll}
    z_{c}^{\ast}&(L_X\ge L_a)\ ,\\
    z_{c}^{\ast}\left(\displaystyle{\frac{L_X}{L_a}}\right)
    ^{\alpha}&(L_X<L_a)\ ,\\
  \end{array}
\right.
\end{equation}
and $p_1$, $p_2$ are given as
\begin{eqnarray}
p_1&=&p^{\ast}_1+\beta_{1}(\log_{10}L_{X}-44)\ ,\\
p_2&=&p^{\ast}_2+\beta_{2}(\log_{10}L_{X}-44)\ .
\end{eqnarray}
Here, the parameters obtained by the fit to X-ray data are (Hasinger,
Miyaji, \& Schmidt 2005): $A_X=6.69\times10^{-7}$ $\rm{Mpc}^{-3}$,
$\log_{10}L^{\ast}_X=43.94\pm0.11$, $\gamma_1=0.87\pm0.10$,
$\gamma_2=2.57\pm0.16$, $z_{c}^{\ast}=1.96\pm0.15$, $\log_{10}L_{a}=44.67$,
$\alpha=0.21\pm0.04$, $p^{\ast}_1=4.7\pm0.3$, $p^{\ast}_2=-1.5\pm0.7$,
$\beta_{1}=0.7\pm0.3$, and $\beta_{2}=0.6\pm0.8$.

\subsubsection{Constraints from the Redshift and Luminosity
Distribution of the EGRET blazars}
\begin{figure}
\epsscale{0.55}
\plotone{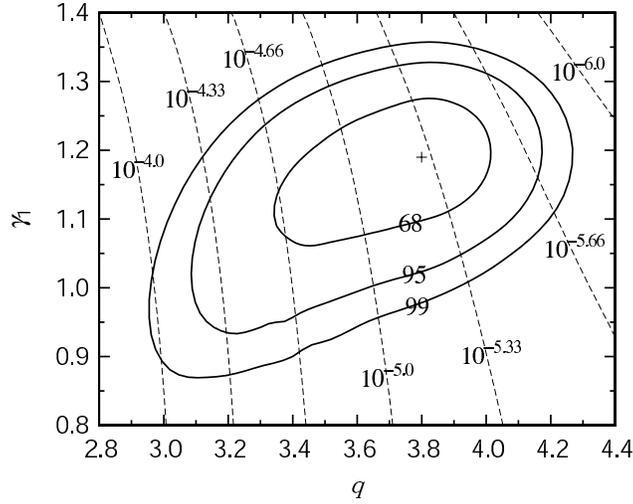}
\caption{The solid contours show the 68\%, 95\% and 99\% C.L. likelihood
contours for the LDDE model parameters [the faint-end slope index
$\gamma_1$ and the gamma-ray to X-ray luminosity ratio,
$q\equiv\log_{10}(L_{\gamma}/L_X)$]. The best-fit values,
$(q,\gamma_1)=(3.80,1.19)$, are shown by the cross. The dashed contours
correspond to $\kappa=10^{-4.0}$, $10^{-4.33}$, $10^{-4.66}$, $10^{-5.0}$,
$10^{-5.33}$, $10^{-5.66}$, and $10^{-6.0}$, respectively, where $\kappa$ is the
normalization ratio of the gamma-ray to soft X-ray luminosity functions.}
\label{fig:CL_contour_LDDE}
\end{figure}

In this model, we take $q$ and $\gamma_1$ as the two free parameters and
fix the rest to the best-fit parameters described in the previous section.
For the radio identification judgement, we use the value of
$\langle p\rangle=3.23$ as obtained from the $L_{\gamma}-L_r$ relation
of the EGRET blazars. The normalization factor $\kappa$ is determined by
fitting the expected total number of blazars to the actually observed
number of the EGRET blazars. In Figure \ref{fig:CL_contour_LDDE} we show
the 68\%, 95\%, and 99\% C.L. contours for the LDDE model (\textit{solid
lines}), with the best-fit parameters $(q,\gamma_1)=(3.80,1.19)$ marked
by \textit{cross}. The best-fit value of $\gamma_1$ is slightly larger
than the value inferred from the SXLF ($\gamma_1=0.87\pm0.10$), but the
SXLF value is within the arrowed region of $\sim95$\% confidence level.
Figures \ref{fig:z_dist_EGRET} and \ref{fig:L_dist_EGRET} show the
redshift and luminosity distributions for the best-fit parameters,
respectively (\textit{solid lines}). It is noteworthy that the LDDE
model can reproduce the redshift and luminosity distributions of the
EGRET blazars better than the PLE model. Quantitatively, the chance
probability of getting the observed deviation estimated from the KS test
is 67.8\% and 99.3\% for the redshift and luminosity distributions, while
these are 3.1\% and 27.0\% for the best-fit PLE model, respectively. These
results indicate that the blazar evolution can better be described by the
LDDE rather than the PLE.

\subsubsection{Blazar Contribution to the Extragalactic Diffuse
Gamma-Ray Background}
\begin{figure*}
\epsscale{1.1}
\plotone{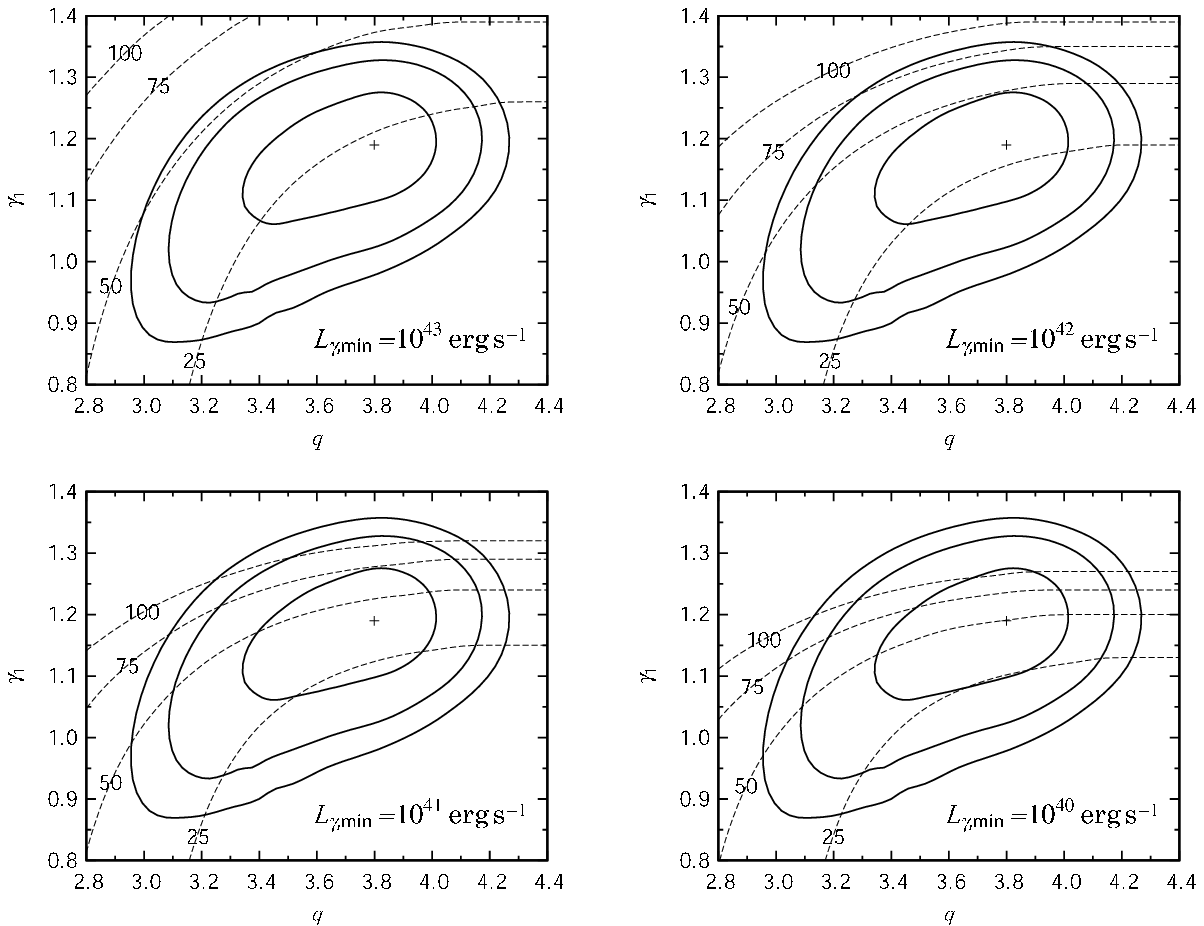}
\caption{The solid contours and crosses are the same as Figure
\ref{fig:CL_contour_LDDE}, showing the fit by the LDDE model. The dashed
contours show 25\%, 50\%, 75\%, and 100\% contribution of unresolved
blazars to the EGRB. The upper left, upper right, lower left, and lower
right panels are for the cases of $L_{\gamma,\rm{min}}=10^{43}$, $10^{42}$,
$10^{41}$, and $10^{40}$ $\rm{erg}$ $\rm{s}^{-1}$, respectively.}
\label{fig:EGRB_LDDE}
\end{figure*}

In Figure \ref{fig:EGRB_LDDE} we present the contours of 25\%, 50\%,
75\%, and 100\% blazar contribution to the EGRB for the LDDE model
(\textit{dashed lines}). We find that unresolved blazars can explain
only 25--50\% of the EGRB for the best-fit parameters. For the case
of $L_{\gamma,\rm{min}}=10^{43}$ erg $\rm{s}^{-1}$, the contour of
100\% blazar contribution is outside the 99\% C.L. region. Still, if
we take the case of $L_{\gamma,\rm{min}}=10^{40}$ erg $\rm{s}^{-1}$,
the LDDE GLF with $\gamma_1\sim1.26$ can marginally explain 100\% of
the EGRB with the parameters within the 68\% C.L. region. However, such
a steep faint-end slope index is not favored from the SXLF
($\gamma_1=0.87\pm0.10$). In the LDDE model the blazar contribution to
the EGRB is strongly dependent on the minimum gamma-ray luminosity of
blazars $L_{\gamma,\rm{min}}$ as well as the faint end slope $\gamma_1$,
since the best-fit faint end slope is $\gamma_1>1.0$.

\section{Predictions for the \textit{GLAST} Mission}
\subsection{Expected Number of \textit{GLAST} Blazars and Their
Contribution to the EGRB}
\begin{figure*}
\epsscale{1.1}
\plotone{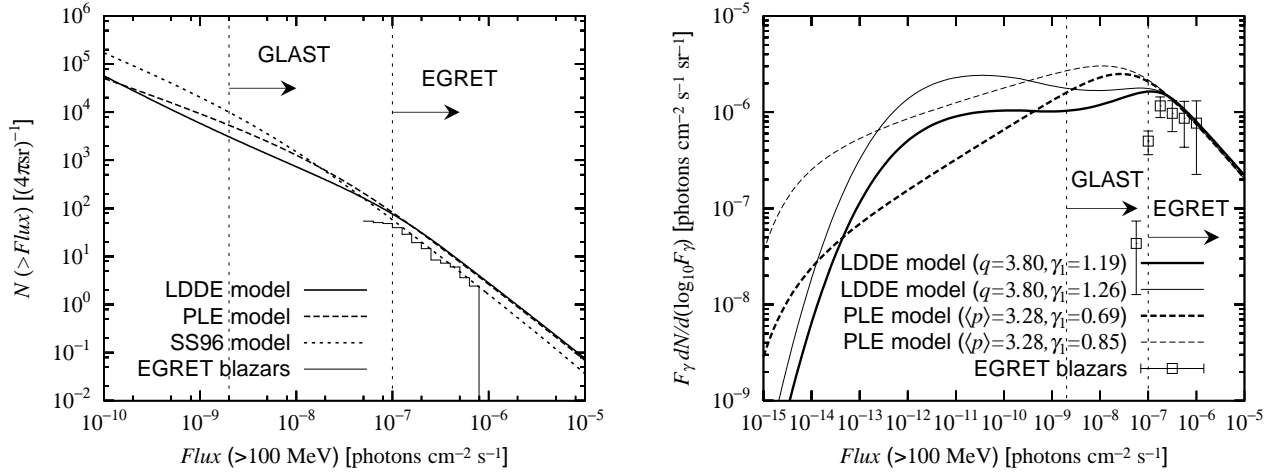}
\caption{Left panel: $\log N-\log F$ distribution (cumulative flux
distribution) of blazars. The solid and dashed curves show the prediction
by the best-fit LDDE and PLE models, respectively. The dotted curve is
derived from the blazar GLF model of SS96. The observed distribution of
the EGRET blazars is shown by the thin solid line. The detection limits
of the EGRET and \textit{GLAST} are also shown in the figure. Right panel:
The same as the left panel, but showing differential flux distribution
multiplied by $F_{\gamma}$, to show the contribution to the EGRB per
logarithmic flux interval. The thick solid and dashed curves are the same
as those in the left panel, but the thin solid and dashed curves show the
LDDE and PLE models with parameters that can explain all the EGRB flux by
unresolved blazars. (See the labels in the panel for the values of the
parameters.)}
\label{fig:count-EGRB}
\end{figure*}

The number of blazars with flux stronger than $F_{\gamma}$ can be
calculated as
\begin{equation}
N(>F_{\gamma})=4\pi\int^{z_\mathrm{max}}_0 dz\ \frac{dV}{dz}
\int^{\infty}_{L_{\gamma}(z,F_{\gamma})}dL_{\gamma}\ 
\rho_{\gamma}(L_{\gamma},z)\ ,
\end{equation}
where $L_{\gamma}(z,F_{\gamma})$ is the gamma-ray luminosity of a blazar
at redshift $z$ whose photon flux above 100 MeV is $F_{\gamma}$ (see eq.
[\ref{eq:Luminosity-Flux}]). In the left panel of Figure
\ref{fig:count-EGRB}, we show the calculated $\log N-\log F_{\gamma}$
relation of blazars. This figure shows that the SS96, our best-fit PLE,
and the LDDE models predict considerably different numbers of blazars
detectable by the \textit{GLAST} ($\sim10000$, 5350 and 3000, respectively),
where we have set the \textit{GLAST} sensitivity limit as
$F_{\lim}=2.0\times10^{-9}$ photons $\rm{cm}^{-2}$ $\rm{s}^{-1}$. Here we
used $L_{\gamma,\min}=10^{40}$ erg $\rm{s}^{-1}$, but the dependence of
the predicted counts on this parameter is small. It is remarkable that the
LDDE model predicts more than three times fewer blazars than the SS96 model.
This is because the LDDE model predicts smaller evolution for less luminous
blazars, and hence paucity of high-$z$ and faint blazars, which have the
dominant contribution to the blazar counts at faint flux in the SS96 or our
best-fit PLE models. This means that we can constrain different blazar GLF
models and their cosmological evolution from the number counts of blazars
detected by the \textit{GLAST}, even without knowing their redshifts. We
also calculate the predicted counts with the LDDE model parameters of
$(q,\gamma_1)=(3.80,1.26)$, which are within the arrowed region of the
68\% C.L. and able to explain 100\% of the EGRB. In this case the prediction
for the \textit{GLAST} is increased to $\sim4700$, but still smaller than
those of the SS96 or the best-fit PLE model.

How much fraction of the EGRB can be resolved by the \textit{GLAST}
mission? To answer to this question, in the right panel of Figure
\ref{fig:count-EGRB} we show the differential flux distribution of
gamma-ray blazars multiplied by flux, showing the contribution to the
EGRB per unit logarithmic flux interval. For the PLE model, we predict
that we will see the peak of the contribution to the EGRB above the
detection limit of the \textit{GLAST} mission, and hence we will resolve
a considerable fraction of the EGRB into blazars, if unresolved blazars
are the major source of the EGRB. The predicted resolvable fraction of
the EGRB flux by blazars detectable by the \textit{GLAST} (but under the
EGRET detection limit) is 33\% and 42\% for the best-fit PLE model and
that with $(\langle p\rangle,\gamma_1)=(3.28,0.85)$, respectively. The
latter model can explain 100\% of the EGRB by unresolved blazars. On the
ohter hand, the LDDE model curves have two peaks of the contribution to
the EGRB as a function of $F_{\gamma}$, because of the complicated nature
of the evolution. We predict that the contribution to the EGRB will
decrease with decreasing flux, just below the EGRET sensitivity limit.
The resolvable fraction of the EGRB by the \textit{GLAST} is 20\% and
26\% for the best-fit LDDE model and that with $(q,\gamma_1)=(3.80,1.26)$,
where the latter model can explain 100\% of the EGRB. As shown in Figure
\ref{fig:count-EGRB}, the dominant contribution to the EGRB comes from
blazars under the \textit{GLAST} detection limit, even if blazars are the
dominant source of the EGRB.

\subsection{Redshift and Luminosity Distribution}
\begin{figure}
\epsscale{0.55}
\plotone{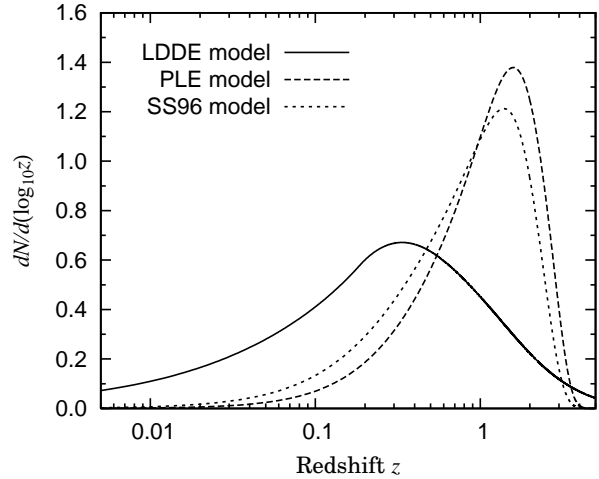}
\caption{Redshift distribution of blazars expected for the \textit{GLAST}
observation. The total number is normalized to the same. The solid and
dashed curves show the prediction by the best-fit LDDE and PLE models,
respectively. The dotted curve is predicted from the blazar GLF model of
SS96. The \textit{GLAST} sensitivity limit is set as
$F_{\rm{lim}}=2.0\times10^{-9}$ photons $\rm{cm}^{-2}$ $\rm{s}^{-1}$.}
\label{fig:z_dist_GLAST}
\end{figure}

\begin{figure}
\epsscale{0.55}
\plotone{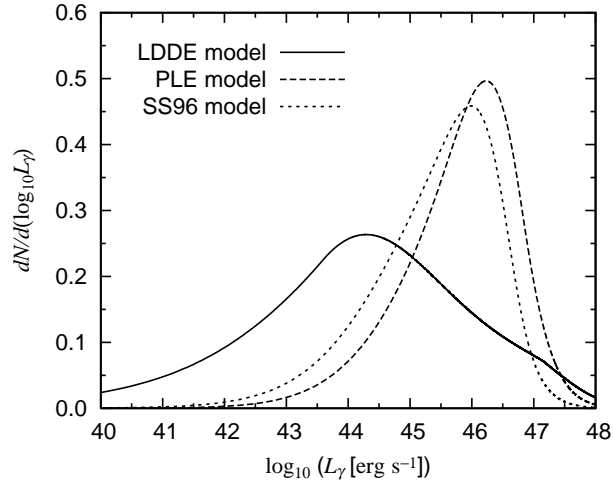}
\caption{Luminosity distribution of blazars expected for the \textit{GLAST}
observation. The total number is normalized to the same. The line markings
are the same as Figure \ref{fig:z_dist_GLAST}. The \textit{GLAST}
sensitivity limit is set as $F_{\rm{lim}}=2.0\times10^{-9}$ photons
$\rm{cm}^{-2}$ $\rm{s}^{-1}$.}
\label{fig:L_dist_GLAST}
\end{figure}

In Figures \ref{fig:z_dist_GLAST} and \ref{fig:L_dist_GLAST} we show the
redshift and luminosity distributions of blazars detectable by the
\textit{GLAST}, respectively. It should be noted that only the shapes of
distribution should be compared, since the total number has been
normalized to the same. It can be seen that the peak of the redshift
distribution in the LDDE model occurs at a lower redshift than the
best-fit PLE or SS96 models. Furthermore, both redshift and luminosity
distributions of the LDDE model are wider than those of the other two
models. Though the redshift must be determined for the future
\textit{GLAST} blazars, this will provide another important information
to discriminate different GLF models.

\section{Discussion}
\subsection{Normalization, Beaming, and Duty Cycle}
\label{section:beaming}
In Figures \ref{fig:CL_contour_PLE} and \ref{fig:CL_contour_LDDE} we
also present the contours of $\eta$ and $\kappa$ (\textit{dashed lines}).
These parameters are possibly related to a beaming effect and/or duty
cycle, and we find $\eta\sim10^{-0.7}$ and $\kappa\sim10^{-5.3}$ for the
best-fit PLE and LDDE models, respectively. The inferred value of $\eta$
is not far from the unity, as expected because the GLF was constructed
from the luminosity function of FSRQs, which are generally believed to be
the same population with blazars. However, the best-fit value
$\eta\sim10^{-0.7}$ is slightly smaller than the unity. In the SSC model,
this value may be explained if not all FSRQs are sufficiently
gamma-ray-loud, or there are more than one components of nonthermal
electrons or emission regions having different beaming patterns (e.g.,
Lindfors et al. 2005). On the other hand, the ERC model may explain this
value of $\eta$ by a single electron component, since the beaming pattern
of the ERC emission is narrower than that of the synchrotron emission or
the SSC emission (Dermer 1995). The observed flux has an angular dependence
$S_{\nu}\propto\mathcal{D}^{2+\alpha}$ for the synchrotron or SSC processes,
and $S_{\nu}\propto \mathcal{D}^{2+2\alpha}$ for the ERC process, where
$\mathcal{D}$ is the Doppler factor and $\alpha$ is the photon spectral
index. Here we define $\theta_{e}$ as the viewing angle of the observer
measured from the jet axis, at which the observed flux is smaller than that
for the direction of the jet axis by a factor of $e$. Then, using the
Lorentz factor $\Gamma=10$ and the typical index ($\alpha_{\gamma}=2.2$,
$\alpha_r=1.0$), we find that $\theta_e\sim3.6^{\circ}$ for the synchrotron
emission, while $\theta_e\sim2.1^{\circ}$ for the ERC emission. Therefore
$\eta\sim(2.1/3.6)^2\sim0.34$ is expected in this case, which is
moderately close to $\eta\sim10^{-0.7}$.

On the other hand, the GLF in the LDDE model is constructed from that of
AGNs selected in the soft X-ray band, most of which are expected to be type
1 AGNs. Since the jet activity is not always observed in AGNs, it is rather
unlikely that all X-ray AGNs have blazar activity. Then, the interpretation
of the small value of $\kappa$ is a combination of the duty cycle $\xi$
(here defined as the fraction of AGNs having the blazar activity) and
beaming of radiation, i.e.,
\begin{eqnarray}
\kappa&=&\frac{\xi}{f_{\rm{type1}}}\ \frac{\Delta\Omega}{4\pi}\\
&=&5\times10^{-6}\left(\frac{f_{\rm{type1}}}{0.2}\right)^{-1}
\left(\frac{\Delta\Omega/(4\pi)}{10^{-3}}\right)
\left(\frac{\xi}{10^{-3}}\right)\ .\quad
\end{eqnarray}
Here, we take the ratio of SXLF to HXLF normalization as the fraction of
type 1 AGNs in all AGNs, $f_{\rm{type1}}$ (Hasinger, Miyaji, \& Schmidt
2005). The beaming expected from the estimated Lorentz factor of blazars
($\Gamma\sim$10--20, e.g., Maraschi \& Tevecchio 2003) is
$\Delta\Omega/(4\pi)\sim1/(4\Gamma^2)\sim10^{-3}$. The small duty cycle
can partially be ascribed to the fraction of radio-loud AGNs to all AGNs,
$f_{\rm{radio}}\sim0.15$ (Urry \& Padovani 1995; Krolik 1999). Comparing
this $f_{\rm{radio}}$ to the inferred $\xi$, it is indicated that about
1\% of radio-loud galaxies have active jets now. It is not unreasonable,
since the jet activity may be sporadic, and radio emission by relic
electrons can be kept for a while after the jet activity ceased.

\subsection{Unidentified EGRET Sources}
Over half of the gamma-ray sources (170 of 271) detected by the EGRET
have not been identified as known astronomical objects. The distribution
of these unidentified sources can be accounted for as the sum of the
Galactic and another isotropic (likely extragalactic) component
(Mukherjee et al. 1995; \"{O}zel \& Thompson 1996). Since most of
the firmly identified extragalactic sources are blazars, it is widely
believed that many of the unidentified extragalactic sources which show
high variability are blazars. Actually, many researchers have been
investigating the unidentified EGRET sources in various wavelengths to
find their counterparts, and some of them, such as 3EG J2016+3657
(Mukherjee et al. 2000; Halpern et al. 2001), 3EG J2006-2321
(Wallace et al. 2002), and 3EG J2027+3429 (Sguera et al. 2004), have
already been identified as blazars. In addition, the number of blazar
candidates in the unidentified sources is now increasing (e.g.,
Sowards-Emmerd, Romani, \& Michelson 2003; Bloom et al. 2004; Wallace,
Bloom, \& Lewis 2005).

These (potential) blazars were initially classified as unidentified
sources likely because of the lack of strong radio emission. Since our
model incorporates the dispersion in radio and gamma-ray luminosities
and selection by radio flux, we can address this issue by estimating how
many ``unidentified blazars'' will be expected due to the lack of
detectable radio flux. We found that this number is 10 and 8 sources at
high Galactic latitude of $|b|>45^\circ$ for the best-fit PLE and LDDE
models, respectively\footnote{The Galactic component of unidentified
sources extends only to $|b|\sim 45^\circ$ (Gehrels et al. 2000).}. In
the third EGRET catalog, there are six low confidence potential blazars
and 19 unidentified sources at $|b|>45^\circ$. The variability of these
unidentified sources have been investigated in some papers (e.g.,
Gehrels et al. 2000; Torres, Pessah, \& Romero 2001; Nolan et al. 2003),
and 5--9 of them are non-variable sources, though definition of
``non-variable'' is different among these papers. Therefore, a substantial
fraction of potential blazars and variable unidentified sources at
$|b|>45^\circ$ can be explained by blazars with radio flux under the
detection limit. It has been suggested that the apparently steady
unidentified sources may be accounted for by forming gamma-ray clusters
(Totani \& Kitayama 2000).

\section{Conclusions}
In this paper, we presented a comprehensive study for the gamma-ray
luminosity function of blazars. We introduced a log-normal distribution
for the ratio of radio to gamma-ray luminosity, and radio detection was
required for the model blazars to become identified sources in the EGRET
catalog. The number of potential blazars and variable unidentified
sources at high Galactic latitude in the EGRET catalog is similar to the
number of ``unidentified blazars'' predicted in our model due to the lack
of detectable radio flux, indicating that our treatment about gamma-ray
blazar identification by radio detection is reasonable. We newly tried
the luminosity dependent density evolution (LDDE) model based on recent
studies of the X-ray luminosity function of AGNs, in addition to the pure
luminosity evolution (PLE) model used in earlier studies.

By performing the maximum likelihood analysis for the redshift and
luminosity distributions of the EGRET blazars, we found that the LDDE model
with the evolutionary parameters inferred from the soft X-ray luminosity
function (SXLF) of AGNs can explain the redshift and luminosity
distributions of the EGRET blazars better than the PLE model with the
evolutionary parameters inferred from the radio luminosity function (RLF)
of flat-spectrum radio-loud quasars (FSRQs). This indicates that blazars
are evolving similarly to type 1 AGNs found in the soft X-ray bands, and
hence the jet activity is universally correlated with the accretion history
of AGNs. We also found that the normalization between blazars and type 1
AGNs is roughly consistent with the unified picture of AGNs, when the
beaming and jet duty cycle are taken into account.

As an implication for the \textit{GLAST} mission, we found that the LDDE
model predicts considerably fewer (by a factor of more than 3) blazars down
to the \textit{GLAST} sensitivity limit, compared with a previous
estimate based on the PLE luminosity function. This can be easily tested
by the mission, giving us important information for the evolutionary
nature of gamma-ray blazars. Redshift and luminosity distributions will
further constrain the different models, though redshift measurements are
required.

Then we examined the contribution of unresolved blazars to the
extragalactic diffuse gamma-ray background (EGRB), which has been
controversial topic in earlier works (100\% in SS96, 25\% in CM98).
We found that only $25$--$50\%$ of the EGRB can be explained with the
best-fit LDDE model, which is similar to the result of CM98 but by
considerably different analysis. On the other hand, according to our
statistical analysis and parameter survey, it is possible to account
for 100\% of the EGRB with a steeper faint-end slope of $\gamma_1\sim1.26$
that is marginally consistent with the arrowed region from the likelihood
analysis of the EGRET blazar distributions. However, such a value is
inconsistent with that inferred from the SXLF of AGNs. Therefore we
conclude that unresolved blazars cannot account for 100\% of the EGRB,
if the jet activity of AGNs is universally correlated to the accretion
luminosity and hence the AGN SXLF is a good description of the blazar
luminosity function and its evolution. It should be noted that the
uncertainty about the extrapolation to high redshift does not change this
conclusion, since almost all of the cosmic X-ray background (CXB) flux can
be explained by the LDDE luminosity function (Ueda et al. 2003). It
indicates that, if the rest of the EGRB is explained by an AGN population,
it must be a different population from EGRET blazars, having different
evolution from X-ray AGNs, and not significantly contributing to the CXB.

Based on the best-fit LDDE model, we predict that the contribution to the
EGRB by blazars will start to decrease with decreasing flux just below the
EGRET sensitivity limit. In the case of the LDDE model with parameters that
can explain 100\% of the EGRB, there are two peaks of the contribution to
the EGRB as a function of flux, and the major contribution comes from
blazars under the \textit{GLAST} detection limit. Therefore it is unlikely
that almost all the EGRB flux is resolved into discrete blazars even by the
sensitivity of \textit{GLAST}.

This work was supported by the Grant-in-Aid for the 21st Century COE
"Center for Diversity and Universality in Physics" from the Ministry of
Education, Culture, Sports, Science and Technology (MEXT) of Japan.

\clearpage

\end{document}